\documentclass[%
 reprint,
showpacs,
 amsmath,amssymb,
 aps,
 pra,
]{revtex4-1}

\usepackage{graphicx}% Include figure files
\usepackage{dcolumn}% Align table columns on decimal point
\usepackage{bm}% bold math
\usepackage{pstricks}

\newcommand{\ket}[1]{|{#1}\rangle}
\newcommand{\bra}[1]{\langle{#1}|}
\newcommand{\prj}[1]{\ket{#1}\bra{#1}}
\newcommand{\ketbra}[2]{\ket{#1}\bra{#2}}
\newcommand{\R}[1]{{\textrm{#1}}}
\newcommand{\cA}{{\mathcal A}}

\newcommand{\cT}{{\mathcal T}}

\newcommand{\Wgeo}{{\mathcal D}}

\newcommand{\deltac}{\delta_{{\rm c}}}

\begin{document}

%\preprint{APS/123-QED}

\title{Cooling the motion of a trapped atom with a cavity field}

\author{Marc Bienert}
 \email{marc.bienert@physik.uni-saarland.de}
\author{Giovanna Morigi}%
\affiliation{AG Theoretische Quantenphysik, Theoretische Physik, Universit\"at des Saarlandes, D-66123 Saarbr\"ucken, Germany
}%


\date{\today}% It is always \today, today,

\begin{abstract}
We theoretically analyze the cooling dynamics of an atom which is tightly trapped inside a high-finesse optical resonator. Cooling is achieved by suitably tailored scattering processes, in which the atomic dipole transition either scatters a cavity photon into the electromagnetic field external to the resonator, or  performs a stimulated emission into the cavity mode, which then dissipates via the cavity mirrors. We identify the parameter regimes in which the atom center-of-mass motion can be cooled into the ground state of the external trap. We predict, in particular, that for high cooperativities interference effects mediated by the atomic transition may lead to higher efficiencies. The dynamics is compared with the cooling dynamics of a trapped atom inside a resonator studied in [Phys. Rev. Lett. {\bf 95}, 143001, (2005)] where the atom, instead of the cavity, is driven by a laser field.
\end{abstract}

\maketitle

%\tableofcontents

\section{Introduction}

Laser cooling exploits the mechanical effects of light in order to prepare atoms and trapped ions at temperatures that vary from microkelvin down to hundreds of nanokelvin \cite{ColloquiaNobel:RMP,Leibfried:2003,Eschner:2003}. The combination with the coupling to the field of high-finesse resonators has brought novel perspectives for the manipulation of the atomic motion: Recent works demonstrated the mechanical effects of light on the atomic motion even at the single-photon level \cite{Pinkse_00,Hood_00,Bushev_PRL04,Ritsch_JOSAB03}, and exploited it in order to achieve trapping \cite{Pinkse_00,Hood_00,Nussman_2005,Puppe_2007} and cooling \cite{Black_PRL2003,Maunz_2004,Nussman_2005,Kimble_PRL2006,Kampschulte_PRL2010,Vuletic_PRL09,Hemmerich_Science12} of the atomic center-of-mass motion.

The coupling to a cavity has been also successfully employed for cooling the motion of an atom  which is trapped by an external potential, such as a harmonic trap  \cite{Kimble_PRL2006,Kampschulte_PRL2010,Vuletic_PRL09}.  In this setting ground-state cooling can be achieved by enhancing the scattering processes which reduce the vibrational excitations of the center-of-mass motion, thus implementing sideband cooling \cite{Leibfried:2003,Eschner:2003,qo:stenholm1986} by means of the strong coupling with the cavity field  \cite{Vuletic_PRA2001}. This implementation extends to trapped atoms the concepts developed for cooling free atoms by means of resonators \cite{Ritsch_JOSAB03,Horak_PRL97,Vuletic_PRL00}. In addition, the quantum nature of the vibrational motion and of the cavity field allows one to identify different physical mechanisms involving photonic and vibrational excitations which contribute to a scattering process and which can thus interfere \cite{Cirac_PRA95,Zippilli_PRL05,bienert:2011}. This interference is a further resource for the manipulation of the atomic motion, since it can be used, for instance, to suppress scattering events leading to heating and/or diffusion of the center-of-mass motion. This knowledge has become particular relevant  in view of recent experiments, which could observe these predictions \cite{Keller_2004,Vuletic_PRL09,Kampschulte_PRL2010,Rempe_CIT}. 

In this article we continue the characterization of laser cooling of trapped atoms inside of resonators by studying the cooling efficiency of a single atom, which is confined inside a high-finesse resonator in the setup sketched in Fig. \ref{fig:setup}, under the condition that the resonator is pumped by a laser. We analyze the temperature and the cooling rate in detail, assuming that the atom is tightly confined and the relevant atomic states compose a two-level dipolar transition which is weakly driven by the cavity field. It is shown that, by suitably tuning the laser parameters, interference processes can enhance the cooling efficiency. The cooling dynamics that are identified are then compared to the ones found in the case in which the atom, instead of the resonator, is driven by a laser and which has been investigated in Refs. \cite{Zippilli_PRL05,Zippilli_PRA05,Zippilli_JMO07}. 

This article is organized as follows. In Sec. \ref{Sec:2} the theoretical model is introduced. The predictions on the cooling efficiency as a function of the laser parameter are presented in Sec. \ref{Sec:3}. The conclusions are drawn in Sec. \ref{Sec:4}. The appendices give some details at the basis of the equations in Sec. \ref{Sec:2} and \ref{Sec:3}.

\section{Theoretical model}
\label{Sec:2}

\begin{figure}
 \includegraphics[width=6.5cm]{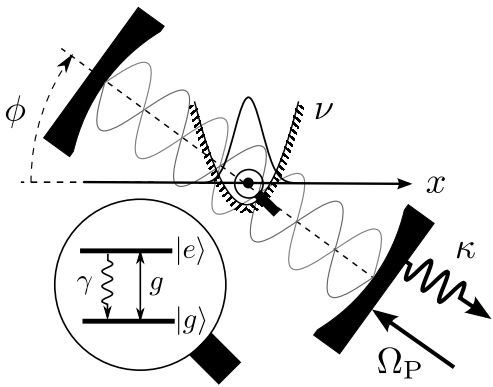}
 \caption{\label{fig:setup} The cooling dynamics of the center-of-mass motion of a trapped atom is studied, under the assumption that the atom is confined within a high-finesse resonator. The motion is cooled by scattering of photons of the cavity mode, which is pumped by a laser.}
\end{figure}

The setup we consider is sketched in Fig.~\ref{fig:setup}. An atom or ion of mass $M$ is confined by a harmonic potential of trap frequency $\nu$, which we here assume to be one dimensional. The internal optical transition couples with a quasi-resonant mode of a high-finesse optical resonator, which is pumped by a laser.  The atomic transition couples the stable electronic state $\ket g$ and the excited state $\ket e$ at transition frequency $\omega_0$ and linewidth $\gamma$, and is quasi-resonant with the cavity mode of frequency $\omega_{\rm c}$. The cavity mode is driven by a laser at frequency $\omega_{\rm P}$ and loses photons at rate $\kappa$. 

The mechanical effects between light and atom are scaled by the recoil frequency $\omega_R=\hbar\omega_0^2/(2Mc^2)$, with $c$ the speed of light, and enter into the dynamics in powers of the Lamb-Dicke parameter $\eta=\sqrt{\omega_R/\nu}$, which is here assumed to be small. Therefore, single-photon absorption and emission events either do not have any mechanical effect or they modify the center-of-mass motion by one quantum of vibration. In this case, the dynamics of the atomic motion is often well described by a set of rate equations which determines the flow of populations $p_m$ between the vibrational levels $\ket m$ of the confining potential. The rate equations have the form \cite{qo:stenholm1986}
\begin{align}
 \frac{d p_m}{d t} = &\eta^2(m+1) A_- p_{m+1} - \eta^2[(m+1)A_+ + m A_-] p_m \nonumber\\ &+ \eta^2m A_+ p_{m-1}\,,
 \label{eq:rateeqcooling}
\end{align}
where $A_-$ ($A_+$) scales the rate of the transition which cools (heats) the atomic motion by one vibrational excitation. From these equations one easily finds the equation governing the dynamics of the mean vibrational number at time $t$, $\langle m\rangle_t=\sum_m mp_m(t)$, which reads 
$\frac{d }{d t}\langle m\rangle_t = -\Gamma \langle m\rangle_t + \eta^2A_+$ with the cooling rate
\begin{align}
 \Gamma = \eta^2\left(A_- - A_+\right)\,. \label{eq:defGamma}
\end{align}
A steady state is found provided that $\Gamma>0$, {\it i.e.} when $A_-$ exceeds $A_+$. In this case, the steady state is a thermal distribution with mean phonon number
\begin{align}
 \langle m\rangle= \frac{A_+}{A_--A_+}\,.
 \label{eq:defm}
\end{align}
The detailed calculation of the cooling rates and of the mean vibrational number at steady-state is performed in the following sections, under the assumption that the coupling of the atom with the cavity field significantly modifies its radiative properties: In this regime the vacuum Rabi frequency $g$ determining the coupling between dipole and cavity field fulfills the relation $g^2/(\kappa\gamma/2)>1$. We will study the cooling dynamics in the so-called good-cavity limit, when $g\gg\kappa$, and in the bad-cavity limit, when $\kappa\gg g\gg \gamma/2$, and compare the results obtained with the studies performed in Refs. \cite{Zippilli_PRL05,Zippilli_PRA05,Zippilli_JMO07}

In the following we will derive the transition rates $A_\pm$ for the considered setup applying the formalism used in Ref. \cite{bienert:2011,Vitali_PRA06}. By evaluating the cooling rate and the mean vibrational number we identify the parameter regions, where cooling is achieved. In these regimes the mean vibrational number at steady state gives the lowest value that can be obtained, and thus one finds whether the parameters allow for ground-state cooling. The cooling rate gives the time scale over which the corresponding distribution can be reached. 

The starting model is the Hamiltonian dynamics of the atom, composed by a two-level transition and the center-of-mass motion, the cavity field mode which couples with the atomic transition in the axial direction. The cavity mode also couples with the modes of the electromagnetic field through the finite mirror transmittivity, determining the finite cavity lifetime, and the transverse modes which couple with the atomic transition, determining its finite radiative lifetime. The Hamiltonian is introduced term by term in the frame rotating at the laser frequency $\omega_{\rm P}$. The effect of the coupling with the modes of the electromagnetic field is modeled using Wigner-Weisskopf theory. 

Before we start, we remark that our analysis restricts to the motion along the $x$ axis, along which the cavity-mode wave vector is assumed to have a finite projection. From a one-dimensional model the efficiency of cooling in three dimensions can be extracted when the Lamb-Dicke regime holds and the trap is harmonic \cite{qo:stenholm1986,Eschner:2003}. In this respect, the results of our analysis can be extended to three dimensions, provided that the diffusion coefficient is properly modified \cite{Morigi:2001}. 

\subsection{Hamiltonian of the atom-cavity system}

We first consider the system composed by atom and cavity mode. Its dynamics are governed by the Hamiltonian
\begin{align}
 H_{\rm sys} = H_{\rm ext} + H_{\rm int} + H_{\rm cav} + W_{\rm P} + W_{\rm C}\,.
\end{align}
The first term on the right-hand side  
\begin{align}
 H_{\rm ext} &=\hbar \nu [b^\dagger b+\tfrac12]
 \label{eq:Hext}
 \end{align}
is the Hamiltonian for the center-of-mass motion in the harmonic trap at frequency $\nu$, with $b$ and $b^\dagger$ the annihilation and creation operators of a vibrational quantum, such that the position and canonically-conjugated momentum of the atom are given by $x = \xi(b+b^\dagger)$ and $p={ i}\hbar/(2\xi)(b^\dagger-b)$, with $\xi = \sqrt{\hbar/(2 M\nu)}$. The Hamiltonian 
\begin{align}
 H_{\rm int} &= -\hbar \delta\prj{e}  \label{eq:Hint}
\end{align}
gives the dynamics of the atomic degrees of freedom in absence of coupling with the e.m. field, with $\delta=\omega_{\rm P}-\omega_0$, and 
 \begin{align}
 H_{\rm cav} &= -\hbar\Delta a^\dagger a
 \label{eq:Hcav}
\end{align}
gives the corresponding term for the cavity field, where  $a$ and $a^\dagger$ are the annihilation and creation operators of a cavity photon and $\Delta=\omega_{\rm P}-\omega_c$ is the detuning between laser and cavity mode. 
The interaction between atom and cavity is described by the term
\begin{align}
 W_{\rm C}&= \hbar g(x)\left[\ketbra{e}{g}a + \ketbra{g}{e} a^\dagger\right]\label{eq:WC}\,.
\end{align}
Here, $g(x)$ is the position-dependent coupling constant
\begin{align}
 g(x) = g \cos(kx\cos\phi  + \varphi)\,,
\end{align}
with $g$ the vacuum Rabi frequency, $k=\omega_c/c$ the wave number of the cavity mode, $\phi$ the angle between wave vector and axis of the motion, while $\varphi$ accounts for the displacement of the trap center with respect to the origin. Moreover, the cavity is driven by a laser, and the corresponding dynamics are described by the operator
\begin{align}
 W_{\rm P} &= \frac{\hbar \Omega_{\rm P}}{2}(a+a^\dagger)\label{eq:WP}
\end{align}
with the strength of the coupling  $\Omega_{\rm P}$, which is related to the input power of the laser $P$ by the equation $\Omega_{\rm P}= 2\sqrt{P\kappa/\hbar\omega_{\rm P}}$.

For later convenience we introduce the single atom cooperativity to quantify the interaction strength. In this paper, we denote it by
\begin{align}
 C= \frac{g^2 \cos^2\varphi}{\kappa\gamma/2}\,,
\label{eq:C}
\end{align}
which accounts for the position dependence of the coupling strength. The parameter $C$ has maximum value $C_0= g^2 /(\kappa\gamma/2)$, which corresponds to the cooperativity when the atom is tightly confined at an antinode of the field. 

Moreover, in order to write the following equations in a compact way 
we introduce the detuning between cavity and atom,
\begin{eqnarray}
\deltac=\omega_c-\omega_0\,,
\end{eqnarray}
such that $\deltac=\delta-\Delta$. 

\subsection{Coupling to the external radiation field}

The total Hamiltonian, including the coupling of the atom-cavity system with the modes of the electromagnetic field external to the cavity, reads
\begin{equation}
\label{H:tot}
H=H_{\rm sys}+H_{\rm emf} +W_{\gamma}+W_\kappa\,,
\end{equation}
where
\begin{align}
H_{\rm emf} =
{\sum_{\vec k, \epsilon}} \hbar[\omega_{\vec k}-\omega_\R{P}] c_{\vec k, \epsilon}^\dagger c_{\vec k, \epsilon}
\end{align}
describes the free dynamics of the electromagnetic field external to the resonator in the reference frame rotating at the laser frequency $\omega_\R{P}$. Here, the sum runs over all modes, identified by the wave vectors $\vec k$ and orthogonal polarization $\epsilon$, the operators $c_{\vec k, \epsilon}$ and $c_{\vec k, \epsilon}^\dagger$ annihilate and create photons in the corresponding mode, and $\omega_{\vec k}=c|\vec k|$ denotes the mode frequency. The operators
\begin{align}
W_{\gamma}&= \hbar{\sum_{\vec k,\epsilon}}^{(\gamma)}
\left[ g_{\vec k,\epsilon}^{(\gamma)} \ketbra{e}{g}e^{i(\vec k\cdot\vec e_x) x}c_{\vec k,\epsilon}+{\rm H.c.}\right]
\label{eq:Wgamma}\,,
\\
W_\kappa &= \hbar {\sum_{\vec k,\epsilon}}^{(\kappa)}
\left[g_{\vec k,\epsilon}^{(\kappa)}  a^\dagger c_{\vec k,\epsilon}+ {\rm H.c.}\right]\
\label{eq:Wkappa}
\end{align}
account for the coupling of the atomic dipole and of the cavity mode with the modes of the external radiation field, respectively. The superscript $\gamma$ refers to the modes whose wave vector is orthogonal to the cavity-mode wave vector. These modes couple with the dipole transition and this coupling gives the radiative instability of the atomic transition at rate $\gamma$, whose explicit form is given by $\gamma=2\pi \sum_{\vec k,\epsilon}^{(\gamma)}|g_{\vec k,\epsilon}^{(\gamma)}|^2\delta(c|\vec k|-\omega_0)$. The superscript $\kappa$ refers to the modes whose wave vector is parallel to the cavity-mode wave vector. These modes couple with the cavity mode via the finite mirror transmittivity and this coupling gives the losses of the cavity at rate $2\kappa$, with  $\kappa=\pi \sum_{\vec k,\epsilon}^{(\kappa)} |g_{\vec k,\epsilon}^{(\kappa)}|^2\delta(c|\vec k|-\omega_c)$ (here we assume that there is no absorption at the mirrors)~\cite{Carmichael}.

\subsection{A tightly confined atom in a weakly-driven cavity}

The complex coupled dynamics of photons and atoms can be reduced to easily solvable rate equations in some specific limits, which we will consider here. These limits are (i) the mean number of photons inside the cavity is very small (which implies that the atomic transition is driven well below saturation) and (ii) the atomic motion is tightly confined by the trap over a spatial region which is much smaller than the light wave length, so that the mechanical effects of light can be treated in perturbation theory. Both situations are realized in several experimental setups. Below we discuss them in some detail.

The assumption that the mean number of intracavity photons $\langle n_c\rangle$ is small corresponds to the inequality
\begin{align}
\langle n_c\rangle \approx \left|\frac{\Omega_{\rm P}/2}{\Delta+i\kappa}\right|^2 \ll 1\,.
\end{align}
Discarding the center-of-mass motion, the relevant state space, in which the cavity and atomic dynamics takes place, is spanned by the states $ \ket{g,0_c}$ and $\ket{g,1_c},\ket{e,0_c}$, whereby the latter two states are coupled by the interaction term, Eq.~\eqref{eq:WC}. When the cavity coupling is strong, it is convenient to consider the basis $ \ket{g,0_c}$ and 
\begin{align}
 \ket{+} &= \cos\theta \ket{e,0_c} - \sin\theta\ket{g,1_c}\,,\nonumber\\
 \ket{-} &= \cos\theta \ket{g,1_c} + \sin\theta\ket{e,0_c}
 \label{eq:dstates}
\end{align}
with $\tan 2\theta = 2 g\cos\varphi/\deltac$ and eigenfrequencies $\omega_\pm$. Equations~\eqref{eq:dstates} give the atom-cavity dressed states \cite{Zippilli_PRA05}.

The considerations made so far were neglecting the mechanical effects of light, considering the atom fixed at position $x=0$. Let us now discuss the coupling with the center-of-mass motion. As we assume that this is tightly confined, we rely that the Lamb-Dicke regime is fulfilled \cite{qo:stenholm1986}. A necessary condition is that the Lamb-Dicke parameter $ \eta = \xi k\ll 1$. If $\eta\sqrt{m}\ll 1$ for all relevantly occupied states $|m\rangle$ of the harmonic oscillator, then we can expand the operators $W_{\rm C}$ and $W_\gamma$ to first order in $x$:
\begin{subequations}
\label{eq:Wexpansion}
\begin{align}
 W_{\rm C} &\approx W_0 + F_{\rm C} x\,,\\
 W_\gamma&\approx W_\gamma^{(0)} + F_\gamma x
\end{align}
\end{subequations}
with $W_0$ and $ W_\gamma^{(0)}$ corresponding to $W_{\rm C}$ and $W_\gamma$, respectively, assuming the atom fixed at $x=0$, and 
\begin{align}
 F_i = \left.\frac{d}{dx} W_i\right|_{x=0}
\end{align}
the gradient of $W_i$ ($i=C,\gamma$) at $x=0$. The operators $F_{\rm C},F_\gamma$ act only on the internal atomic and cavity degrees of freedom and have the dimension of a force \cite{Cirac_PRA92,Morigi_PRA03}. 

\subsection{Light scattering at the atom-cavity system}

\begin{figure}
 \includegraphics[width=8cm]{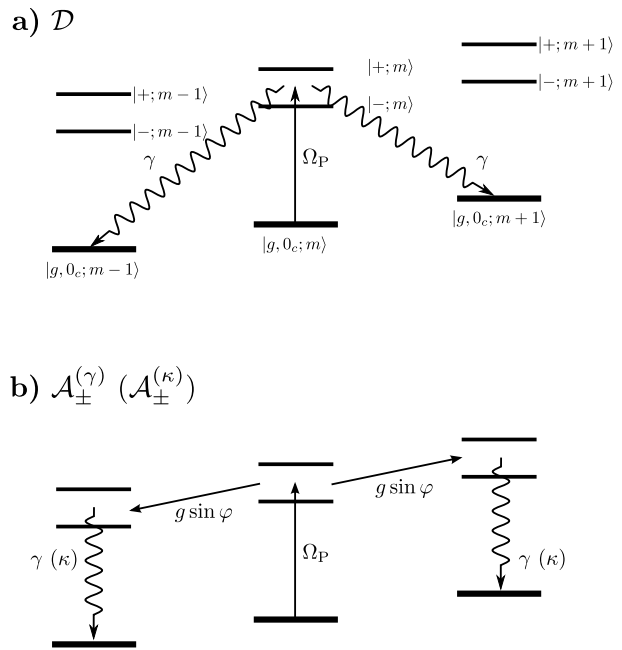}
 \caption{\label{fig:processes} Basic processes of photon scattering leading to the contributions $\Wgeo$, (a), $\cA^{(\gamma)}_\pm$ and $\cA^{(\kappa)}_\pm$ (b), in Eq.~\eqref{eq:Apm}. The processes start from state $|g,0_c,m\rangle$ and connect it to states $|g,0_c,m\pm 1\rangle$ via a scattering path whose intermediate states are the dressed states  $|\pm,m\rangle$ and $|\pm,m\pm 1\rangle$. The straight arrows indicate coherent processes, the wiggled arrows the incoherent processes. One notes that there is no direct excitation $|g,0_c,m\rangle\to |\pm,m\pm 1\rangle$, differing from the case in which the atom is driven by the laser \cite{Zippilli_PRA05,Zippilli_JMO07}. This property leads to a major difference between the cooling dynamics of the two situations.}
\end{figure}

The Lamb-Dicke regime and the assumption of weakly-driven cavity allow us to determine the rate of photon scattering by means of perturbation theory in the small parameters $\eta$ and $\epsilon=|\Omega_{\rm P}/2(\Delta+i\kappa)|$. Using standard methods (a summary of the relevant steps is found in Appendix \ref{App:A}) we evaluate the heating and cooling rates entering rate equation \eqref{eq:rateeqcooling}. Analogously with the case in which the driving laser directly couples with the atom \cite{Zippilli_PRL05,Zippilli_PRA05,Zippilli_JMO07}, they can be cast in the form  
\begin{align}
 A_\pm= \Wgeo + \cos^2\phi \sin^2\varphi\left\{\cA^{(\gamma)}_\pm + \cA_\pm^{(\kappa)}\right\}\,, 
 \label{eq:Apm}
\end{align}
which is the sum of three positive terms. Here, $\Wgeo$ results from the diffusive motion the atom undergoes due the photon recoil of spontaneous emission in lowest, non-vanishing order in $\eta$. The corresponding process is depicted in Fig.~\ref{fig:processes}a). The second term on the right-hand side, $A_\pm^{(\gamma)}$, is due to  processes, where the atom-cavity system absorbs a photon from the pump, then undergoes a cavity-induced transition which changes the motional state, and finally emits a photon due to spontaneous emission in zeroth order in $\eta$. The processes connected to $A_\pm^{(\kappa)}$ are similar, except that the system relaxes due to cavity losses, see Fig.~\ref{fig:processes}b). Depending on whether the ratio $\gamma/\kappa$ is small or large, $A_\pm^{(\kappa)}$ or $A_\pm^{(\gamma)}$ can dominate one over the other. Equation \eqref{eq:Apm} also shows explicitly that the contribution of these two terms can be large when (i) the axis of motion is aligned with the cavity, so that the mechanical effect due to absorption or emission of a cavity photon is maximal, and (ii) when the atom is close to a node of the cavity mode, where the gradient of the electric field, and thus the exerted force, is maximal.  The necessary condition for cooling to occur is thus that $\sin\varphi\neq 0$, namely, the atom shall not be localized at an antinode of the standing wave. The other obvious necessary condition is that $\cos\phi\neq 0$, corresponding to a non-vanishing projection of the cavity mode wave vector on the motional axis, and thus to a non-vanishing recoil associated with the absorption and emission of cavity photons \cite{Footnote:1}. 

The diffusion rate $\Wgeo$ can be recast in the form $$\Wgeo=A_0D_0\,,$$ where $D_0$ is of order unity and depends on the geometry of the setup and the properties of the considered atomic transition, while $A_0=\gamma\rho_e\cos^2\varphi$ is the rate at which cavity photons are scattered by the atom into the modes external to the cavity. Here, $\rho_e\cos^2\varphi$ is the stationary occupation of the excited state at zero order in the Lamb-Dicke parameter with
\begin{align}
 \rho_e =\frac{\Omega_{\rm P}^2}{4}\frac{g^2}{f(\Delta,\delta_c)}\,.
 \label{eq:A0}
\end{align}
The denominator of this equation is the function
\begin{align}
 f(\Delta,\delta_c) &= \left[\Delta(\delta_c+\Delta)-\frac{\kappa\gamma}{2}(1+C)\right]^2\nonumber\\&+ \left[\Delta\frac{\gamma}{2}+(\delta_c+\Delta)\kappa\right]^2\,,
\label{eq:f}
\end{align}
whose real poles are the resonances of the atom-cavity system. We note that the coefficient vanish when the atom is tightly confined at the node of the cavity standing wave: In this case, in the Lamb-Dicke regime diffusion processes are suppressed \cite{Cirac_PRA92,Zippilli_PRA05}.

The other rates take the form
\begin{align}
 \cA_\pm^{(\gamma)} = \gamma\frac{\rho_e}{f(\Delta\mp\nu,\delta_c)}&\Bigg\{\left[(\Delta\mp\nu)\delta -\frac{\kappa\gamma}{2}(1 -C)\right]^2 \nonumber\\&+\left[(\Delta\mp\nu)\frac{\gamma}{2}+\delta\kappa\right]^2\Bigg\}\,, \label{eq:Apmgamma}\\
 \cA_\pm^{(\kappa)} =  2\kappa\frac{4\rho_e\,g^2\cos^2(\varphi)}{f(\Delta\mp\nu,\delta_c)}&\left[(\delta\mp\nu/2)^2+\frac{\gamma^2}{4}\right]\,.
 \label{eq:Apmkappa}
\end{align}
They contain the function $f$, but evaluated at $\Delta\mp\nu$: their behavior is thus dominated by the poles of $f$ shifted by $\pm\nu$, corresponding to the red and blue sidebands of the corresponding resonance. We also note that, differing from the rates $\cA_\pm^{(\gamma)}$, the rates $\cA_\pm^{(\kappa)}$ vanish also at the nodes of the standing wave, since in this case there is no mechanical process that couples states $|g,0_c,m\rangle$ and $|g,0_c,m\pm 1\rangle$ by emission of a cavity photon. 

The rates are expressed in terms of the cooperativity $C$, Eq. \eqref{eq:C}: This term explicitly highlights the cavity QED effects on the motion of the atom. We note that the limit $C\ll 1$, where the cavity is a small perturbation to the atomic dynamics, gives the rate equations of a two-level transition in free space \cite{Cirac_PRA92}. In this case, ground-state cooling is achieved in the strong confinement limit, $\gamma\ll\nu$, where sideband cooling can be implemented, as shown in Appendix \ref{App:B}.  

In the next section we discuss the results obtained when $C\ge 1$ and provide some insight into the dynamics. In order to identify the parameter regimes where the atomic motion can be efficiently cooled we characterize the poles of function $f$. The requirement that the first term of $f(\Delta+\nu)$ in Eq.~\eqref{eq:f} vanishes leads to equation
\begin{align}
 \delta = \frac{\frac{\kappa\gamma}{2}(1+C)}{\Delta+\nu}-\nu\,.
 \label{eq:coolres}
\end{align}
This equation has the same form as the one found for the case in which the atom is driven: see Eq. (55) of Ref. \cite{Zippilli_PRA05}. The only difference is that the role of atomic and cavity detuning are here interchanged. This is an obvious consequence of the fact that the cavity field drives the transition $|g,0_c\rangle \to |g,1_c\rangle$ (which is then coupled to $|e,0_c\rangle$ by the cavity field), while a transverse laser drives the transition $|g,0_c\rangle\to |e,0_c\rangle$ (which is then coupled to $|g,1_c\rangle$ by the cavity field). Substantial differences between the two cases however arise when considering interference effects in the coherent processes leading to a mechanical excitation, as we will discuss in the next section. 

\section{Results}
\label {Sec:3}

We now analyze the predictions of the rate equations, focusing on the regime where they allow for a steady state solution, and thus cooling of the motion by radiative scattering. The cooling dynamics are characterized by contour plots of the mean vibrational number at steady state $\langle m\rangle$, Eq. \eqref{eq:defm}, and of the cooling rate $\Gamma$, Eq. \eqref{eq:defGamma}, as a function of the detuning $\delta=\omega_{\rm P}-\omega_0$ between pump and atom and of the detuning $\Delta=\omega_{\rm P}-\omega_c$ between pump and cavity mode. 

In the following we will consider the regime in which $C\ge 1$. Our attention will especially focus on the regime in which the cooling rate can be enhanced by means of resonances, considering regimes where transitions leading to diffusion and heating can be suppressed by quantum interference.  

\subsection{Good-cavity limit}
\label{sec:coolSCS}

\begin{figure}
 \includegraphics[width=\columnwidth]{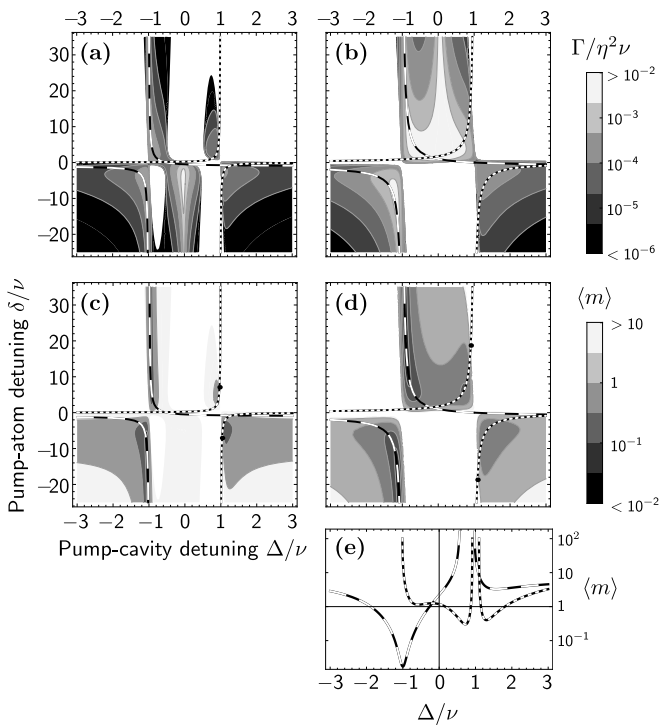}
\caption{\label{fig:gamma}  
Cooling rate $\Gamma$, Eq. \eqref{eq:defGamma}, and mean phonon number $\langle m\rangle$, Eq. \eqref{eq:defm}, as a function of the detunings $\delta$ and $\Delta$ (in units of the trap frequency $\nu$) and in the good-cavity limit, assuming $\gamma\gg\kappa$. The parameters are $\gamma=10\nu$, $\kappa=0.025\nu$, cooperativity $C=3$ (left panel) and $C=15$ (right panel). White areas denote parameter regions where the motion is heated, the dashed curve in the contour plots indicates Eq.~\eqref{eq:coolres}, the dotted curves Eq.~\eqref{eq:qicond}. The bottom plot displays $\langle m\rangle$ along the dashed and dotted curves for $C=15$. The other parameters are $\Omega_{\rm P}=0.01\nu$, $\Wgeo=1$, $\phi=0$, $\varphi = 0.45\pi$.}
\end{figure}

We first analyze the good-cavity limit, taking $\gamma\gg \kappa$. In this case spontaneous emission dominates over cavity decay. Figure \ref{fig:gamma} displays the contour plots of the cooling rate, Eq.~\eqref{eq:defGamma}, and of the mean vibrational number, Eq.  \eqref{eq:defm}, as a function of the detunings $\Delta$ and $\delta$. The resonance condition, Eq.~\eqref{eq:coolres}, which gives the position of the red sideband of the dressed states, is indicated in the plot by the dashed line~\cite{footnote:4}. Parameter regions where the cooling rate $\Gamma$ becomes negative are marked by white areas: For these parameters no steady state can be found for the equations of motion, indicating heating in the limit of validity of the theory. 

From the figures one readily sees that relatively large cooling rates and low vibrational occupations are found along the curve corresponding to Eq. \eqref{eq:coolres}. This behavior is more pronounced when the cooperativity is increased. 

For comparison, in the limit in which the coupling with the resonator can be treated classically (corresponding to small cooperativity, see App.~\ref{App:B}, and large intracavity photon numbers), the atom is cooled on the red side of the pump-atom detuning, i.e., for $\delta<0$: Optimal temperatures and cooling rate would correspond to the Doppler cooling limit, see \cite{qo:stenholm1986}. This situation is indeed recovered inside the cavity but for large values of $|\Delta|$ and small values of $|\delta|$, namely, when the pump is far-off resonance from the cavity mode but almost resonant with the atomic transition frequency. In this case the dynamics reduce to an effective two-level system with ground state $|g,0_c\rangle$ and excited state $|e,0_c\rangle$, while state $|g,1_c\rangle$ is only virtually occupied since far-off resonance. The cooling rate is therefore small and scales at large $|\Delta|$ as $1/|\Delta|^2$, while the linewidth of the excited state determines the lowest vibrational occupation one can achieve. For the choice made here, where $\gamma>\nu$, we obtain Doppler cooling by setting $\delta=-\gamma/2$ with $\langle m\rangle\propto \gamma/4\nu>1$, see Eq.~\eqref{eq:meanmDoppler}. 

\begin{figure}
 \includegraphics[width=\columnwidth]{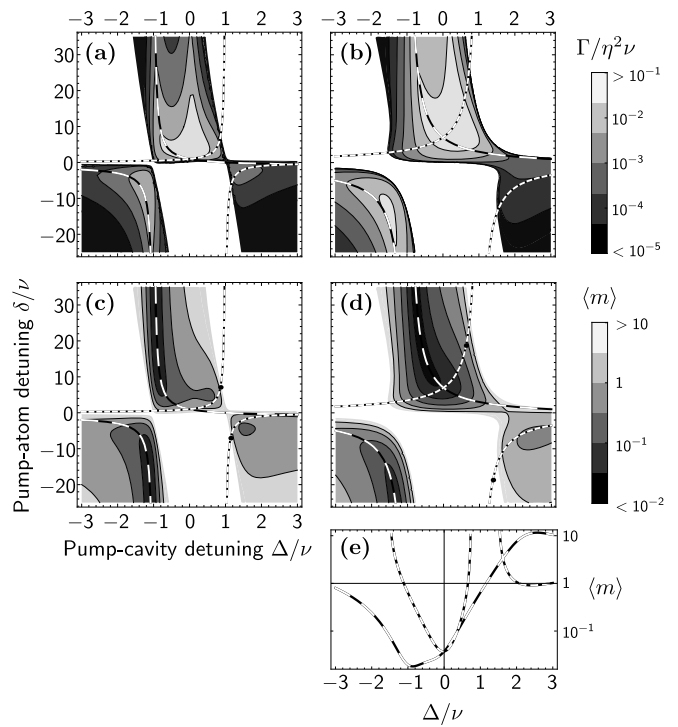}
\caption{\label{fig:gammab}  
Same as Fig.~\ref{fig:gamma}, except that $\kappa = 0.1\nu$.}
\end{figure}

The other asymptotic behavior is identified for large values of $|\delta|$ and small values of $|\Delta|$, namely, when the pump is far-off resonance from the atomic transition but almost resonant with the cavity mode. In this limit the dynamics reduce to an effective two-level system composed by the ground state $|g,0_c\rangle$ and a dressed state which mostly coincides with $|g,1_c\rangle$ with a small component of state $|e,0_c\rangle$ whose occupation scales with $\sin^2\theta$, see Eq. \eqref{eq:dstates}. In leading order the rates $\cA_\pm^{(\gamma)}$ then contribute to the diffusional motion of the atom only, whereas cooling takes place due to light scattering via the cavity. The linewidth of the relevant dressed state is close to the linewidth $\kappa$ of the resonator, and for $\kappa<\nu$, the optimal choice of parameters corresponds to setting $\Delta=-\nu$, thereby performing sideband cooling. The minimal vibrational occupation scales with $\langle m\rangle_\infty\sim \kappa^2/(4\nu^2)+O(1/C)$, while the cooling rate decreases with $\sin^2\theta$.

These two limiting cases are similar with the ones found when the atom, inside the resonator, is driven by the laser, under the condition that the role of the incoherent processes is interchanged: $\kappa< \gamma/2$. This latter situation has been analyzed in Ref. \cite{Zippilli_JMO07}. In general, the case here studied and the one analyzed in \cite{Zippilli_JMO07} share several analogies, provided that in the setup where the atom is driven, the laser is perpendicular to the axis of the motion, and hence there are no mechanical effects associated with the absorption of a laser photon (see Fig. 3(c) and (d) of Ref. \cite{Zippilli_JMO07}). Nevertheless, some fundamental differences can be identified that, in the case here considered, are due to the diffusion associated with spontaneous decay (which is strongly suppressed when the major source of losses is cavity decay as in \cite{Zippilli_JMO07}). 

Further parameter regions where cooling can be efficient are found when the numerator of $\cA_+^{(\gamma)}$, Eq.~\eqref{eq:Apmgamma} becomes small. This is generally fulfilled when the first term in the curly brackets of Eq.~\eqref{eq:Apmgamma} vanishes \cite{footnote:2}, corresponding to the condition
\begin{align}
 \delta = \frac{\frac{\kappa \gamma}{2}(1-C)}{\Delta -\nu}\,.
 \label{eq:qicond}
\end{align}
Curve ~\eqref{eq:qicond} is indicated in Fig. \ref{fig:gamma} by the dotted curves. This curve extends to $\Delta=\nu$ and explains the rectangular cooling area for $\delta>0$. The cooling rate $\Gamma$ in this area becomes largest when the pump drives the cavity close to resonance. Some insight into the phenomenon can be gained by identifying the excitation paths leading to interference, which are depicted in Fig.~\ref{fig:processes}b): There are generally four pathways to connect the initial and final state via the two pairs of dressed states $\ket{\pm;m}$ and $\ket{\pm; m+1}$. Under condition~\eqref{eq:qicond} their probability amplitudes add up in a way that $\cA_+^{(\gamma)}$ is reduced.

Figure \ref{fig:gamma}(e) displays the mean phonon number along the dashed and broken lines in the contour plot at largest cooperativity. The dotted line, along which this quantum interference effect is strongest, exhibits two minima around $\Delta=\nu$ at moderate temperatures. Similar processes are also found in the corresponding case when the atom is driven. In Fig.~\ref{fig:gammab} we plotted the cooling rate and mean phonon number for larger values of $\kappa$. In this case the signatures of quantum interference for the cooling process are less pronounced, but still distinguishable. For $C=15$, the lowest temperature along the dashed curve is then found around $\Delta =0$, where the laser is at the same time resonant with the red sideband of a dressed state, see Fig.~\ref{fig:gammab}(d) and (e). At lower cooperativity $C=3$, the interference effect manifests itself by the two cooling region around $\Delta = \nu$, as shown in Fig.~\ref{fig:gammab}(c).

\subsection{Bad-cavity limit}

\begin{figure}
 \includegraphics[width=\columnwidth]{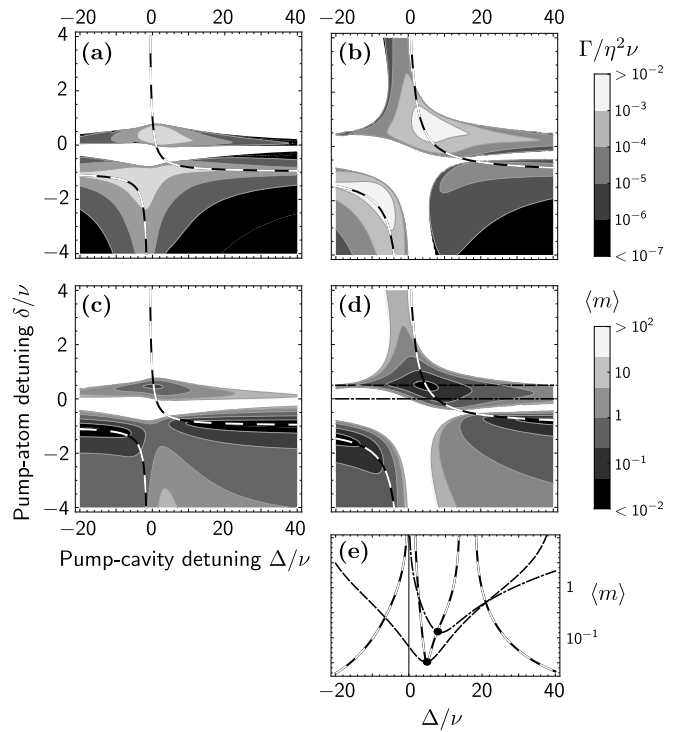}
 \caption{\label{fig:kappa} Same as Fig. \ref{fig:gamma} but in the bad-cavity limit, $\kappa\gg\gamma$. The parameters are $\gamma=0.15\nu$, $\kappa=4.5\nu$, $C=5$ (left panel) and $C=25$ (right panel). The other parameters are $\Omega_{\rm P}=0.1\nu$, $\Wgeo=1$, $\varphi=\pi/3$ and $\phi=0$.}
\end{figure}

We now discuss the case when cavity losses dominate over spontaneous decay, {\it i.e.} $\kappa\gg\gamma$. Figure \ref{fig:kappa} displays the contour plots for the mean vibrational number and cooling rate. We first observe that the cooling regions stretch well beyond the parameter regime along the curve in Eq. \eqref{eq:coolres}, which is associated with the red sideband of the dressed-state resonances. This behavior becomes more enhanced when the cooperativity increases, as comparison between left and right panel shows. In addition, along the curve given by Eq. \eqref{eq:coolres} one observes also heating regions. These behaviors can be understood in terms of interference processes, some of which we will discuss in the following. 

In order to better understand the dynamics, we will analyze the transition rates $\cA_\pm^{(\kappa)}$, Eq.~\eqref{eq:Apmkappa}, which determine the rates $A_\pm$ when $\kappa\gg\gamma$. By setting $\gamma=0$ they read
\begin{align}
 A_\pm = &
 \cos^2\phi\rho_e\, \times\nonumber\\
 & 2\kappa\frac{g^2\sin^2 \varphi \cos^2\varphi\left(2\delta\mp\nu \right)^2}{[(\Delta\mp\nu)(\delta\mp\nu)-g^2 \cos^2\varphi]^2+(\delta\mp\nu)^2\kappa^2}.
 \label{eq:Apmbadcavity}
\end{align}
Lead by the results in Fig. \ref{fig:kappa}, below we study their behavior in order to better understand the cooling dynamics. 

\subsubsection{Pump resonant with the atom}

We first focus on the region about $\delta=0$, where the pump is resonant with the atomic transition. In a certain interval of values of the detuning $\Delta$, this region exhibits cooling, which reaches almost unit ground state occupation with relatively fast cooling rates for values of $\Delta$ between 0 and 20$\nu$. More efficient cooling is found when the resonance condition~\eqref{eq:coolres}  is simultaneously fulfilled. 

In order to get some insight into this behavior, we set $\delta=0$ in Eq. \eqref{eq:Apmbadcavity} and find
\begin{align}
 A_\pm= \cos^2\phi \rho_{e} \, 2\kappa\frac{\nu^2 g^2\sin^2\varphi\cos^2\varphi}{\kappa^2\nu^2 + \left[\nu(\nu\mp\Delta)-g^2\cos^2\varphi\right]^2}\,.
\label{eq:eit}
\end{align}
In this form, one recovers the form of the rates found in EIT-cooling \cite{Morigi_PRL00,Morigi_PRA03}, where the atom is in free space and the relevant transitions form a $\Lambda$-shaped configuration \cite{footnote:3}. The mechanism at the basis of the observed behavior is indeed similar, since the states $|g,0_c\rangle$, $g,1_c\rangle$, and $|e,0_c\rangle$ form a $\Lambda$ transition, where a quasi-dark state can be found since the intermediate state $|g,1_c\rangle$ decays faster than  $|e,0_c\rangle$, see \cite{Rice1996}. While in EIT cooling the carrier excitation of the atom is suppressed by destructive interference of the excitation paths of the electron, likewise, for the situation considered here, only transitions connected by mechanical interaction can populate the state $\ket{g,1_c}$. In contrast to EIT cooling, the interference effect does not cancel atomic diffusion resulting from spontaneous emission: For small, but finite $\gamma$ diffusion is still present, but cavity relaxation takes place on a faster time scale. Minimal temperatures are found at the pump-cavity detuning $\Delta = g^2\cos^2\varphi/\nu - \nu$, for which from Eq.~\eqref{eq:eit} we find 
\begin{align}
 \langle m\rangle &= \left(\frac{\kappa}{2\Delta}\right)^2\,,\\
 \Gamma &\approx 2\tilde{\eta}^2 \rho_{e}\frac{g^2\cos^2\varphi}\kappa
\end{align}
with $\tilde{\eta}=\eta\cos\phi\sin\varphi$. We remark that this interference effect is observed when the cavity is weakly driven \cite{Rice1996}. An analogous effect is observed in the good cavity limit when the atom is driven and the laser is resonant with the cavity mode \cite{Rice1996,Vuletic_Science11}. In those settings it can be exploited for performing ground state cooling thereby suppressing the excited state occupation, and therefore diffusion \cite{Zippilli_PRL05,Zippilli_PRA05}. 

\subsubsection{Quantum interference at $\delta=\nu/2$}

Quantum interference in the scattering process  gives rise to the cooling regions at non-negative values of $\delta$. The cooling region extends over a large region of the plot and is most efficient when condition Eq.~\eqref{eq:coolres} is fulfilled. A maximum is found about the value $\delta\simeq \nu/2$. An analysis of Eq.~\eqref{eq:Apmbadcavity} show that at this value, indeed, heating transitions can be suppressed when the radiative decay rate $\gamma$ is sufficiently small. In second order in the Lamb-Dicke expansion, the temperature and cooling rate take the form
\begin{align}
 \langle m\rangle &\approx \frac{1}{C}\frac{9}{16}\left[1+\frac{\cot^2\varphi}{\cos^2\phi}\Wgeo\right]\,,\\
 \Gamma &\approx \tilde{\eta}^2\rho_e \frac{32}{9} \frac{g^2\cos^2\varphi}{\kappa}\,,
\end{align}
which have been here evaluated for the detuning $\Delta = 2g^2\cos^2\varphi/3\nu-\nu$, which maximizes the ratio $A_-/A_+$. 

\begin{figure}
 \includegraphics[width=\columnwidth]{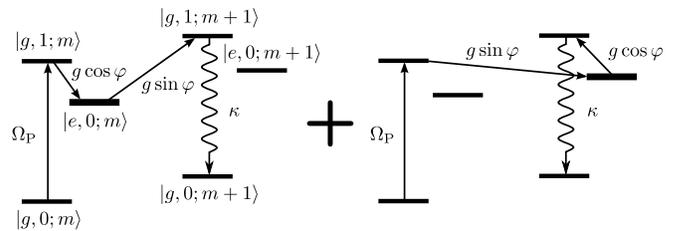}
 \caption{\label{fig:suppheatproc} 
Excitation paths contributing to the heating transition $|g,0_c,m\rangle\to |g,0_c,m+1\rangle$, here depicted in the lowest order of the Born expansion in the cavity-atom coupling $g$ and for $\kappa\gg\gamma$. The amplitude of each path scales with $1/(\Delta+\delta_c)$ (left) and $1/(\Delta+\delta_c-\nu)$ (right). For $\delta=\Delta+\delta_{\rm c}=\nu/2$ the energy defect has opposite sign and the two processes interfere destructively.}
\end{figure}

The interference effect leading to the suppression of heating does not depend on the coupling strength $g$, as Eq.~\eqref{eq:Apmkappa} suggests. This allows one to investigate the underlying scattering processes in lowest non-vanishing order in $g$. The corresponding processes are sketched in Fig.~\ref{fig:suppheatproc}: Two excitation paths lead to destructive interference for heating transitions when $\delta = \nu/2$ (and ideally $\gamma=0$). This picture is easily extended to arbitrary values of $g$. We also note that a similar behavior was identified for the case in which the atom is driven, but in the good-cavity limit \cite{Zippilli_PRL05}. 

Figure~\ref{fig:kappa}(e) displays the mean vibrational occupation number $\langle m\rangle$ along the cuts indicated by the corresponding curves in the contour plot \ref{fig:kappa}(d). The lowest temperature can be achieved for large detunings $\Delta$, where cooling at the red sideband of the narrow dressed state, which here almost corresponds to $\ket{e,0_c}$, can be realized. The two minima of the temperature (black bullets) correspond to the cooling mechanisms which are due to quantum interference at optimal detunings: The minimum at $\Delta\approx 5\nu$ corresponds to the case where destructive interference suppresses heating, whereas at $\Delta\approx 8\nu$ cavity-induced transparency governs the cooling dynamics.

\section{Conclusions}
\label{Sec:4}

Cooling of a trapped atom has been analyzed, under the condition that the atomic dipolar transition strongly couples to the mode of a high-finesse cavity, which is pumped by a laser. The parameter regimes have been identified for which ground state cooling can be performed. We predict that, for large cooperativities, ground-state cooling can be performed with the help of quantum interference between the motion and the cavity photonic excitations, which can suppress undesired scattering processes. 

The cooling dynamics has been compared with the case where the atom is directly driven by the external laser, which has been extensively studied in Refs.  \cite{Zippilli_PRL05,Zippilli_PRA05,Zippilli_JMO07}. The analysis of the equations show that in most parameter regimes the two situations can be mapped one into the other by swapping the atomic and the cavity variables. The major difference between the two cases is due to (i) the mechanical effect of spontaneous emission, which makes the incoherent process due to the radiative instability of the atom not equivalent to cavity losses, and (ii) the mechanical effect of the laser on the atom, which makes this dynamical processes not mappable to the case in which the cavity is driven. Indeed, when the system is pumped by a transverse laser coupling to the atom, interference between the cavity and the laser can suppress atomic excitation, and therefore diffusion. In addition, the mechanical effect of the laser can interfere with the ones of the cavity field, thereby providing further control tools for suppressing heating transitions. In this respect, the configuration in which the atom is driven offers richer dynamics than the case of the driven cavity.  

The cooling dynamics identified in this work could be observed in current experimental setups~\cite{Keller_2004,Kampschulte_PRL2010, Rempe_CIT,Vuletic_PRL09}. Information about the temperature and cooling rate can be accessed by means of electron shelving techniques \cite{Leibfried:2003,Eschner:2003} and by measuring the spectrum of the light at the cavity output \cite{Bienert07}. 

The concepts here analyzed can also be extended for developing efficient cooling schemes of molecules, for molecular transitions like the ones considered in Ref. \cite{Kowalewski}, and optomechanical systems \cite{Aspelmeyer:2011}. Our results may find applications in quantum technology such as quantum sensors  \cite{Aspelmeyer:2011} and atom-photon interfaces in quantum networks, where the position of the atom in the field needs to be controlled with nanoscale precision \cite{Kimble_2007,Ritter_2012}.

\begin{acknowledgments}
We acknowledge discussions with T. Kampschulte, D. Meschede, and S. Zippilli. This work has been partially supported by the European Commission (Integrating Project AQUTE; STREP PICC), by the Ministerio de Ciencia e Innovaci{\'o}n (QOIT Consolider-Ingenio 2010; FIS2007-66944; EUROQUAM ``CMMC''), by the BMBF ``QuORep'', and by the German Research Foundation (DFG).
\end{acknowledgments}

\begin{appendix}

\section{Evaluation of the scattering rates}
\label{App:A}

We use scattering theory in order to evaluate the rates $A_\pm$ which determine the cooling. This approach allows for a direct identification of the underlying physical processes. We split the total Hamiltonian, Eq.~\eqref{H:tot}, according to $ H = H_0 + V$, where
\begin{align}
 H_0 &= H_{\rm ext} + H_{\rm int} + H_{\rm cav} + H_{\rm emf} + W_0
 \label{eq:H0}
\end{align}
and 
\begin{align}
 V =W_\gamma^{(0)} + W_\kappa + W_{\rm P} + x (F_{\rm C} + F_\gamma).
\end{align}
%In Eq.~\eqref{eq:H0} the contributions of the internal atomic and cavity degrees of freedom are merged in the Hamiltonian $H_{\rm opt}$.
The rates of photon scattering, which change the motional state of the atom by one vibrational quantum are given by
\begin{subequations}
\label{eq:ApmDef}
\begin{align}
 (m+1) A_+ &= \frac{2\pi}{\hbar} \sum_{\rm f} |\cT_{{\rm f}_+}|^2\delta(E_{\rm i} - E_{\rm f})\,,\\
 m A_- &= \frac{2\pi}{\hbar}\sum_{\rm f}|\cT_{{\rm f}_-}|^2\delta(E_{\rm i} - E_{\rm f})
\end{align}
\end{subequations}
and contain the sum over all transition matrix elements
\begin{align}
 \cT_{{\rm f}_\pm} = \lim_{\varepsilon\to 0}\, \bra{{\rm f}_\pm} V G(E_{\rm i}+i\varepsilon) W_{\rm P}\ket{{\rm i}}
 \label{eq:transampl}
\end{align}
with final states $\ket{f_\pm}$. The initial and final states of the scattering process are 
\begin{align}
 \ket{\rm i} &= \ket{g, 0_c; m; 0_{\vec k,\epsilon}} \label{eq:istate}\,,\\
 \ket{\rm f_\pm} &= \ket{g, 0_c; m\pm 1;  1_{\vec k,\epsilon}}\,.
\end{align}
The initial state is stable in zeroth order in $\eta$ and $\Omega_{\rm P}$. The atomic motion is initially in state $\ket{m}$, with the external field in the vacuum. After a scattering process which changes the vibrational motion, the number of motional excitation in the final state is changed by a single quantum. The sum over all final states in Eq.~\eqref{eq:ApmDef} is thus a sum over all possible modes of the scattered photon, whereby the Dirac $\delta$-function in Eqs.~\eqref{eq:ApmDef} ensures energy conservation between the energy of the initial state, $E_{\rm i} = (m+\frac12)\hbar\nu$, and of the final state $E_{\rm f}$. The evolution operator is found from the resolvent $G(z)$, which  we approximate in lowest order in $\Omega_{\rm P}$ and up to first order in the Lamb-Dicke parameter $\eta$:
\begin{align}
 G(z) &= \frac{1}{z - H}\nonumber\\
 &\approx G_0(z) + G_0(z) x (F_\gamma + F_{\rm C}) G_0(z)\,.
\label{eq:G}
\end{align}
Here
\begin{align}
 G_0(z) = \frac{1}{z - H^{\rm eff}_0}
\label{eq:G0}
\end{align}
contains the non-Hermitian operator 
\begin{align}
H^{\rm eff}_0 = H_0 - i\hbar \left[\frac{\gamma}{2}\prj e +\kappa a^\dagger a\right].
\label{eq:H0eff}
\end{align}
The evolution operator for $t>0$, and hence the transition amplitudes, is found for $z=E+ i\varepsilon$ with $\varepsilon\to 0_+$, see Eq.~\eqref{eq:transampl}. 

With the help of Eqs.~\eqref{eq:transampl}-\eqref{eq:G0} it is straightforward to evaluate the rates $A_\pm$, Eq.~\eqref{eq:ApmDef} up to first non-vanishing order in the small parameters. Details about the calculation can be found in Refs.~\cite{Vitali_PRA06,bienert:2011}. Reference \cite{bienert:2011}, in particular, contains the model considered here as a limiting case.

\section{Rate equations for small cooperativity}
\label{App:B}

In the limit of small cooperativity, $C\ll 1$, the rates in Eqs. \eqref{eq:Apm} are approximated by the expressions
\begin{align}
 A_\pm = A_0\left[D_0 + \cos^2\phi\tan^2\varphi\frac{\delta^2 + \gamma^2/4}{[\delta \mp \nu]^2+\gamma^2/4}\right]\,,
\label{eq:ApmlowC}
\end{align}
which has been obtained in leading order in the small parameter $C$. We note that the contribution of scattering processes, where the photon is emitted by the cavity, is here negligible: In fact rates $A_\pm^{(\kappa)}$ are of order $C$ with respect to rates $A_\pm^{(\gamma)}$. The rates~\eqref{eq:ApmlowC} coincide with the ones found for cooling an atom in free space, whose two-level transition is driven by a standing-wave field  with Rabi-frequency $\Omega_{\rm SW}/2 = g |\epsilon|$. This situation has been extensively discussed in \cite{Cirac_PRA92}. 

Ground-state cooling is found for $\gamma\ll\nu$. This is the sideband cooling limit. Setting the detuning $\delta =-\nu$, the mean vibrational number and cooling rate read
\begin{align}
 \langle m\rangle &\approx \left(\frac{\gamma}{4\nu}\right)^2\left[1+4\frac{\cot^2\varphi}{\cos^2\phi}D_0\right]\,,\\ 
 \Gamma & \approx \eta^2 \sin^2\varphi \cos^2\phi\frac{\Omega_{\rm SW}^2}{\gamma}\,.
\end{align}
In the weak coupling limit, $\gamma>\nu$, the minimum temperature is achieved for $\delta = -\gamma/2$, leading to a final vibrational occupation number and cooling rate
\begin{align}
 \langle m\rangle &= \frac{\gamma}{4\nu}\left[1+\frac{\cot^2\varphi}{\cos^2\phi}D_0\right]-\frac12\label{eq:meanmDoppler}\,,\\
 \Gamma  &= \eta^2 \sin^2\varphi \cos^2\phi\frac{2 \nu \Omega_{\rm SW}^2}{\gamma^2}
\end{align}
corresponding to the Doppler cooling limit. 

This limit differs from the situation encountered when the atom instead of the cavity is driven: Then, the effect of the cavity field  at leading order is completely negligible, see \cite{Zippilli_PRA05}.

It is interesting to note that when the vacuum Rabi frequency is large but the atom is trapped at a node of the cavity field and hence $C=0$, the cooling dynamics are very similar to the free space case \cite{Cirac_PRA92}.

\end{appendix}

%\bibliography{article,qo,qobooks,publist}% Produces the bibliography via BibTeX.

\begin{thebibliography}{99}
\bibitem{Footnote:1}
We note  that the conditions on the geometry are significantly different when the laser drives the atom. In that case the atomic motion can be cooled by photon scattering into the cavity mode even if the mechanical effect of the cavity field is zero and/or the atom is trapped at an antinode of the cavity standing wave, see \cite{Zippilli_PRA05}.
\bibitem{footnote:2}
For $\Delta = \mp\kappa\sqrt{C-1}+\nu$, $\delta=\pm\frac{\gamma}{2}\sqrt{C-1}$ and $C>1$ the rate of heating $\cA_+^{(\gamma)}=0$ even vanishes completely. However, this choice of detunings implies $\cA_+^{(\kappa)}=1$, and heating effects due to cavity dissipation dominate. In Figs.~\ref{fig:gamma} and \ref{fig:gammab} these values are marked by dark dots.
\bibitem{footnote:3}
The mapping can be performed when the Rabi frequency of the strong laser driving one branch of the $\Lambda$-system is replaced by $g\cos\varphi$, and the atomic linewidth of the upper state is substituted by the cavity linewidth.
\end{thebibliography}

\end{document}